# Multi-time small-area estimation of oil and gas production capacities by Bayesian multilevel modeling

Hiroaki Minato
Office of Energy Production, Conversion & Delivery
U.S. Energy Information Administration (EIA)

1 January 2025

## Abstract

This paper presents a Bayesian multilevel modeling approach for estimating well-level oil and gas production capacities across small geographic areas over multiple time periods.  Focusing on a basin, which is a geologically and economically distinguishable drilling region, we model the production capacities of its wells grouped by area and time.  Regularizing our inferences with priors, we model area-level and time-level variations as well as well-level variations, incorporating lateral length, water usage, and sand usage at each well.  The Maidenhead Coordinate System is used to define uniform geographic areas, many of which contain only a small number of wells in a given time period.  First, a Bayesian small-area model is built, using data from the Bakken region from February 2012 to June 2024.  Then, the model is expanded to contain temporal dynamics in the production capacities.  In addition to general time components, water and sand usage intensities are modeled in estimating production capabilities over time.  We find the Bayesian multilevel modeling approach provides a flexible and robust framework for modeling and estimating oil and gas production capacities at area and time levels and for informing area-time predictions with uncertainties.

## 1. The initial model—a small-area estimation model

The oil and gas industry continues to play a significant role in the economy.  Developments in the U. S. oil and gas industry, particularly those associated with oil and gas extraction or fracking, continue to have large impacts on the U. S. and international markets.  There are demands to understand what is going on in the oil and gas fields and what to expect in the future.  This methodology working paper is to accompany or supplement the substantive working paper on the topic, which focuses on the data acquisition and preparation as well as the economic and technological analyses.  Our unified goals are to: provide framework to understand variations and reduce uncertainties; provide data and statistics to corroborate, quantify, or challenge untested claims; facilitate deeper discussions and analyses in a larger community; and support the openness of data, methods, algorithms, and codes for public use.

We first describe the Bayesian multilevel modeling and estimation of well-level oil and gas production *capacities* in multiple areas in a specific time period. By capacity, we mean a latent quantity that represents a well's expected production potential, and it is our main estimand.

A basin (or "region") is a geography of interest, as the basin is an economically established drilling area and has a similar geology. We assume existing or potential wells in the same basin have a similar level of oil and gas production capacities.

However, we expect there is a substantively significant variation in the production capacity among the wells even in the same basin. This well-level variation is decomposed in our multilevel model by grouping the wells by the so-called MHB—the 8.5-to-10-square-mile Maidenhead block that is based on the Maidenhead Coordinate System (Lieskovsky and Zyren, 2022). Within each MHB $b$, the well-level variability in the production capacity is linearly modeled with the lateral length $L_i$, i.e., an MHB-level simple linear regression of the observed production capacity $Y$ on $L$:

$$Y_i \sim \text{normal}(\alpha_{b[i]} + \beta_{b[i]} L_i, \sigma_Y), \ i = 1, \dots, N, \quad (1)$$

where $N$ is the number of wells in the basin, $b[i]$ indicates MHB $b$ that the well $i$ belongs to, and $\sigma_Y$ is the standard deviation of $Y$ or the residual parameter of the multilevel model. This model probabilistically permits $Y$ to be negative, though $Y$ is theoretically non-negative. However, we think it is easy to work with and can well approximate the unknown true data generation mechanism.

Intuitively, the lateral length $L$ represents a line segment, not an end point. That is, the production capacity of a well with $L = l$ is an accumulation or integration of the production capacity of the well from 0 to $l$ on the lateral line. So, the slope $\beta$ of the regression model is interpreted as the area-level mean production capacity of the well along the lateral line. Roughly, we imagine a solid cylinder of oil or/and gas with the circular area $\beta$, being laid down over the length $l$ with the volume $\beta \times l$.

We also have the water and sand usage measurements at each well. In theory, these two quantities are highly correlated (collinear) because water is used to carry sand. And, both water and sand are used to extract oil or/and gas so they are non-negatively associated with the actual production of oil or/and gas.

For simplicity and robustness, we first modeled the mean of $\beta_b$ as a linear function of the area-level average water usage $\overline{W}_b$. Specifically, we regressed $\beta$ on $\overline{W}$ over $b$:

$$\beta_b \sim \text{normal}(\gamma + \delta \overline{W}_b, \sigma_\beta), \ b = 1, \dots, B, \quad (2)$$

where $B$ is the number of MHB's in the basin. (We also tried constraining each $\beta_b$ to be non-negative; however, this led to some sampling/convergence problems with the current models and data.)

When the number of observed wells $n_b$ is small in a block $b$, we call the block a small area and estimating, for example, $\text{E}(Y|\text{MHB} = b)$ becomes a small-area estimation problem.

In this demonstration, we use a multisource dataset complied from publicly available data for Bakken region from 21 February 2012 to 12 June 2024[1].  The number of wells $N$ is 3,848 and the number of MHB's $B$ is 415.

## Notes on Bayesian multilevel modeling

Multilevel modeling is a "generalization of regression methods, and as such can be used for a variety of purposes, including prediction, data reduction, and causal inference from experiments and observational studies" (Gelman, 2006).  A regression model can be generalized by adding more predictors at the individual and group levels and by allowing the slope and the intercept to vary by group.  Thus, a focus of multilevel modeling is to quantify sources of variation and to share information across different groups (populations, scenarios, time frames, datasets, and so on).  Compared to no pooling (which tends to overfit) or complete pooling (which tends to underfit), partial pooling by multilevel modeling often gives more reasonable inferences or more accurate predictions, especially when predicting group averages.  "One intriguing feature of multilevel models is their ability to separately estimate the predictive effects of an individual predictor and its group-level mean, which are sometimes interpreted as 'direct' and 'contextual' effects of the predictor" (Gelman, 2006).  Further, the Bayesian approach we take allows "inclusion of prior information (which can also be viewed as regularization or stabilization of inferences)" (Gelman et al., 2020).  Even when the number of observations per group is small and thus a least squares regression with group indicators would give unacceptably noisy estimates, a Bayesian multilevel regression can partially pool varying coefficients in a regularized way for more stable predictions and better generalizability and transportability.

When we say a prior is "weakly informative", we mean what Gelman et al. (2020) mean: "If there's a reasonably large amount of data, the likelihood will dominate, and the prior will not be important.  If the data are weak, though, this 'weakly informative prior' will strongly influence the posterior inference.  The phrase 'weakly informative' is implicitly in comparison to a default flat prior."

## Transformation of variables

Examining the data, we have made the following decisions.  First, we have chosen the oil production variable as our outcome variable $Y$.  The oil and gas productions from a given well are highly dependent, though the dependency is not easy to characterize.  Since the oil data seem "cleaner" and distributed more "nicely", we are using the observed oil productions as proxies for the latent oil and gas production capacities.

Considering the positivity of values and the shape of the distributions, any of the original variables (the oil production variable, the lateral length variable, and the water usage variable) can be log-transformed.  It could affect the computation and would also change the meaning of the models, while reshaping the data distributions.  For the initial model, no log-transformations were applied.

Meanwhile, each of the variables was standardized by subtracting its average and dividing it by its standard deviation.  This step makes the regression coefficients comparable (in a multiple regression model or among multiple models of the same form), but it, more importantly, constrains the scales of the priors and the likelihood.

---

[1] As mentioned earlier, another EIA working paper on the topic is in progress and will describe the source data and their compilation processes as well as the substantive backgrounds of this paper.

Rescaling could be done by using 2 times the standard deviation, which leads to the standard deviation of 0.5 after rescaling. When one has a binary predictor in the model, Gelman (2008) suggests the latter rescaling for comparability as the standard deviation of a Bernoulli random variable with the parameter $p$ is $\sqrt{p(1-p)} \leq 0.5$. This rescaling is implemented in the later models in this paper.

## Priors and (in)sensitivity

All the priors were specified as the "generic" weak prior distribution. That is, $\alpha_b$, $\gamma$, and $\delta$ were assumed to be independently and normally distributed with the mean 0 and the standard deviation 1, and $\sigma_Y$ and $\sigma_\beta$ independently and half-normally distributed with the mean 0 and the standard deviation 1. Weaker prior distributions normal(0, 10) and half-normal(0, 10) were also tried but did not affect the posterior inferences very much in this example.

## MCMC/NUTS convergence is fast and good

We implemented the model using Stan in R (RStan) through RStudio (Stan Development Team, 2024). The no-U-turn sampling (NUTS) algorithm, an adaptive variant of the Hamiltonian Monte Carlo method, produced samples that converged very quickly and stably to the posterior distribution (three chains of 500 warm-up cases and 2,500 sample cases in each chain converged in a few minutes). So, this was a good sign for our model being reasonable at least for the given data[2].

Cross-validation is often used to "estimate how well a model would predict previously unseen data by using fits of the model to a subset of the data to predict the rest of the data" (Vehtari, 2024). For example, the LOO package in R (Vehtari et al., 2024) computes approximate leave-one-out cross-validation (LOO-CV) for a fitted model, using Pareto smoothed importance sampling (PSIS). However, we focus here on estimation rather than prediction so we do not pursue cross-validation.

## Posterior predictive check and RMS of deviation seem reasonable

After fitting the model, we predicted the outcomes using the posterior distributions of the model parameters. The discrepancy between the predicted values based on the posterior means (in red) and the observed values (in blue) was large and systematic, namely, the predicted was less skewed to the right or more symmetric than the data. Thus, some of the small and large values were not well captured by the current model.

---

[2] Andrew Gelman's "Folk Theorem of Statistical Computing" says, "When you have computational problems, often there's a problem with your model" (https://statmodeling.stat.columbia.edu/2008/05/13/the_folk_theore/).

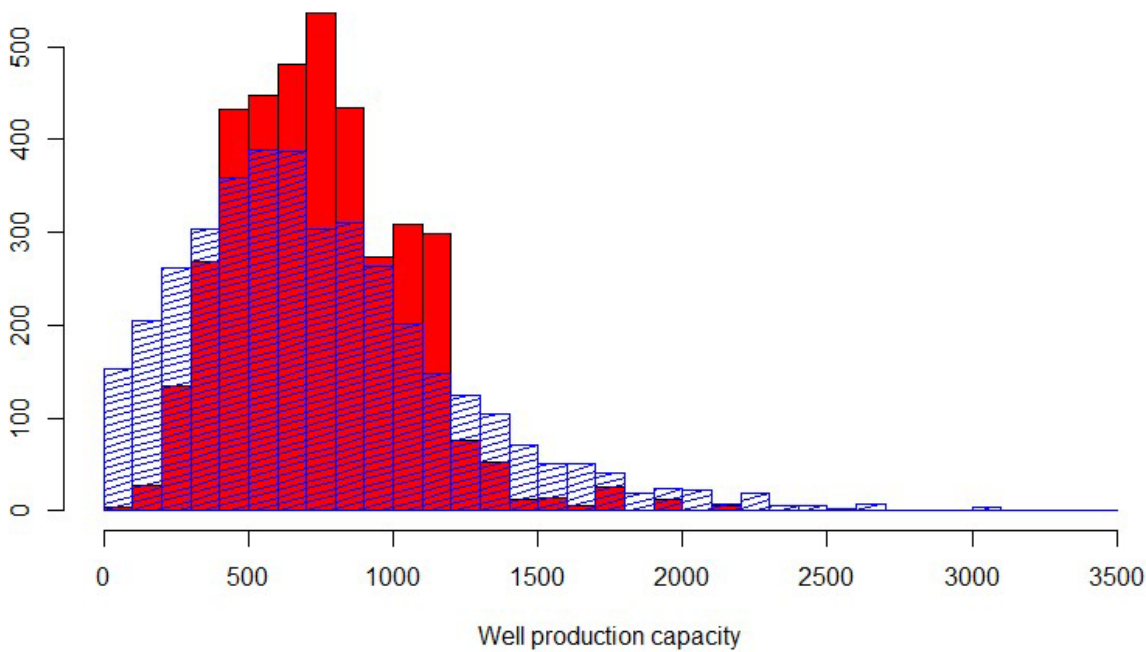

Computing the root-mean-square deviation (RMSD) between the predicted and the observed, we got about 344. About 90 % of the discrepancies were distributed between -525 and 594.

### What's next?

At this point, we can try improving the model fit by (1) combining or multivariate modeling the oil and gas production variables, (2) modeling the spatial correlations among the blocks by a joint prior with some covariance structure, (3) partially pooling the blocks for the intercept $\gamma$ and the water parameter $\delta$, (4) incorporating any external data such as the geological survey data, (5) further experimenting with the likelihood and the priors, and so on.

We also think it is important to validate our general modeling approach or "software" with the data from other years or/and basins, i.e., to check the generalizability of our model.

However, we will take on another important dimension—time—in the next section.

## 2. Multi-time expansion of the small-area estimation model

In Section 1, we built a Bayesian multilevel model of well-level oil production capacity in areas (MHB's), small or otherwise, of a given basin (Bakken) in a certain time period (from 21 February 2012 to 12 June 2024).

Now, we take into account the time structure defined by a calendar year-month period and expand the model by introducing the time-effect parameters $\tau_t$ ($t = 1, \ldots T$) and making $\beta_b$ additionally time-dependent as $\beta_{b,t}$ ($b = 1, \ldots, B$ and $t = 1, \ldots T$) in (1):

$$Y_i \sim \text{normal}(\alpha_{b[i]} + \tau_{t[i]} + \beta_{b[i],t[i]} L_i, \sigma_Y), \quad i = 1, \ldots, N, \quad (3)$$

where $N$ is the total number of unique wells among $B$ MHB's in the basin over $T$ times. $\beta_{b,t}$ remain unconstrained. Some of the wells could become non-operational over time while some new (or even old) wells could come into operation over time. Thus, instead of some longitudinal or time-series data structure, we define a stack of multiple cross-sectional datasets with a discrete time variable as a grouping variable. Note that implicitly there are $T$ time-indicator variables—one for each time—just as there are $B$ area-indicator variables.

Further, instead of (2), we now specify a deterministic function for $\beta_{b,t}$ in (3):

$$\beta_{b,t} = \gamma_t + \delta_b \bar{E}_b, \quad b = 1, \ldots, B \text{ and } t = 1, \ldots, T, \quad (4)$$

where $\bar{E}_b$ is the average water-sand intensity in each MHB $b$. Ignoring vertical lengths of wells as insignificant effects on the oil and gas production capacities, the water-sand intensity $E_i$ is defined for each well $i$ as:

$$E_i = \frac{W_i + S_i}{L_i},$$

where $W_i$ is the water usage in gallons, $S_i$ is the sand usage in pounds, and $L_i$ is the lateral length in feet at well $i$. (If $L_i = 0$, we can treat the value as a "bad" measurement and remove the case from the modeling or impute the value as if missing. Or, we might just set $E_i = 0$. In the current data, $L_i > 0$ for all cases.) Then, $\bar{E}_b$ is simply:

$$\bar{E}_b = \sum_{i \in b} E_i / N_b,$$

where $N_b$ is the number of wells in MHB $b$.

We interpret that $\delta_b \bar{E}_b$ approximates the level of technology or innovation associated with the combined water and sand usage efficiency level in MHB $b$, which is assumed to be invariant over time. That is, the level of technology or innovation in MHB $b$ is some function of $\bar{E}_b$, which does not depend on the time, and is linearly approximated by $\delta_b \bar{E}_b$.

We now have the following multilevel model with multiple areas and times:

$$Y_i \sim \text{normal}(\alpha_{b[i]} + \tau_{t[i]} + (\gamma_{t[i]} + \delta_{b[i]} \bar{E}_{b[i]}) L_i, \sigma_Y), \quad i = 1, \ldots, N. \quad (5)$$

Or, equivalently,

$$Y_i \sim \text{normal}([\alpha_{b[i]} + \delta_{b[i]} \bar{E}_{b[i]} L_i] + [\tau_{t[i]} + \gamma_{t[i]} L_i], \sigma_Y), \quad i = 1, \ldots, N. \quad (6)$$

The data $Y_i$, $E_i$ (before averaging), and $L_i$ (not $L_i$ within $E_i$) are again standardized—with their observed averages and standard deviations. However, we use 2 standard deviations in the denominators so that the standardized variables get 0.5's as their standard deviations (Gelman, 2008). This makes those continuous variables comparable with the group indicator variables MHB and Time in terms of scaling. (However, this is not essential in the current problem; indeed, it was not done for the initial model.)

The priors in (3) and (4) are all normal distributions with spatial independence, temporal dependence, and slightly stronger regularization (i.e., the standard deviations of 0.5's are used, compared to 1's in the initial model—however, recall that the data are standardized to get the standard deviations of 0.5's):

$$\alpha_b \sim \text{normal}(0, 0.5), \qquad b = 1, \ldots, B,$$

$$\tau_1 \sim \text{normal}(0, 0.5), \qquad \tau_t \sim \text{normal}(\tau_{t-1}, 0.5), \qquad t = 2, \ldots, T,$$

$$\gamma_1 \sim \text{normal}(0, 0.5), \qquad \gamma_t \sim \text{normal}(\gamma_{t-1}, 0.5), \qquad t = 2, \ldots, T,$$

$$\delta_b \sim \text{normal}(0, 0.5), \qquad b = 1 \ldots, B,$$

and

$$\sigma_Y \sim \text{normal}^+(0, 0.5).$$

Observations in our stacked cross-sectional dataset are not independent, because some wells appear more than once in the data over the monthly time periods. Although the same well is likely to have different values of $W$, $S$, and $L$ whose effects on $Y$ could be dominating at a given time, some of the variation of $Y$ could depend on the time. Thus, in the above, we built the autocorrelated priors for the time parameters $\tau_t$ and $\gamma_t$ ($t = 1, \ldots, T$).

We have run the model on the horizontal wells with non-negative values of $Y_i$ (oil production), $W_i$, and $S_i$ and positive values of $L_i$ in the Baken region. (It is possible to remove wells with "extreme" values, considering their data quality issues or modeling difficulties.) With the sample size $N$ = 3,848, the number of areas/blocks $B$ = 415, and the number of times/years $T$ = 10 (from 2015 to 2024), RStan's NUTS required only a couple of minutes to produce three chains of 3,000 samples (including 500 warm-up samples) in each chain without divergences or R-hat values above 1.1 (Vehtari et al., 2019).

We interpret $\gamma_{t[i]} + \delta_{b[i]} \bar{E}_{b[i]}$ as the production/extraction efficiency (per lateral foot) of a "typical" well $i$ at the time $t$ and in the MHB $b$ that has the area-level average water-sand intensity $\bar{E}_b$. The expected production capacity of the well $i$ is this area-and-time-level production efficiency times its lateral length $L_i$ plus the base area-effect $\alpha_{b[i]}$ and the base time-effect $\tau_{t[i]}$—i.e., the mean in (5).

Note that with the original variables the intercepts are not forced to be zero, as we are not interested in modeling where the input variables are zeros. Besides, the linearity is only an approximation, and zero intercepts would reduce the freedom in the linear approximation. Thus, we included $\alpha_b$ and $\tau_t$ as independent parameters from $\delta_b$ and $\gamma_t$, respectively.

Also, we didn't constrain $\gamma_{t[i]} + \delta_{b[i]} \bar{E}_{b[i]}$ or $\delta_{b[i]}$ to be non-negative in our model fitting. For some $(t, b)$ or some $b$, it is possible for those parameters to be negative, when production capacities are on average smaller for wells with longer lateral lengths. In those situations, we could either accept any negative estimates of $Y_i$ (when de-standardized) or bound them below by zero as assumptions or expectations.

The model now depends on $T$. We cannot predict $Y$ at the $(T + 1)$th time based on the model that is built on the data from the last $T$ times, just as we cannot predict $Y$ at some external block $(B + 1)$ from the model built from the $B$ blocks. In fact, since $Y$ is observed whenever $L$, $W$, and $S$ are observed, there is nothing to predict at the well level.

One might be interested in estimating the underlying parameters such as the mean oil/gas production capacity of block $b$ at time $t$: $\mu_Y(b, t) = \mathrm{E}(Y|MHB = b, Time = t)$. (In the first section, the mean oil/gas production capacity of block $b$ over time was estimated: $\mu_Y(b, \{1, \dots T\}) = \mathrm{E}(Y|MHB = b, Time \in \{1, \dots T\})$.) When the number of observations is small in block $b$ at time $t$, the simple average of $Y$ in $b$ at $t$ would not be very precise (or has a large variance)—a small area estimation problem. Also, since wells are not randomly drilled in a block at a time and may not "represent" all potential wells in the block at the time, the simple average may be biased in estimating the block-time parameter for $Y$ such as $\mu_Y(b, t)$.

We model a regression function and its coefficients themselves and estimate the parameters from data. Further, by the Bayesian approach, we utilize our prior information or uncertainty also to regularize the estimation. We know the variance problem would be well addressed almost by construction. However, the bias problem, if any, is currently ignored (though it could be modeled). For example, we could assume every observed well is a higher-performing well in the block at the time—i.e., selection biases. As every small area-time estimate would be pulled to the area-time regression prediction, the bias in the higher-performing wells would be reduced, while the bias in the lower-performing wells would be increased.

One way out of this bias problem may be to define the population of wells in block $b$ at time $t$ to be a collection of higher-performing wells (however they may be defined), not of all possible wells in $b$ at $t$. That is, we change the domain of our construct $Y$. This definition or assumption may not be bad, as it is actually more realistic than assuming the drilled and operated wells are randomly located.

In any case, a premise of small-group estimation is that you can learn some aspects of a specific individual, area, or time from other individuals, areas, or times. They together form a sample, from which some inferences are drawn about the population of interest. Bayesian multilevel modeling is a natural and powerful method for small-group estimation problems[3].

## Posterior predictive check and RMS of deviation seem reasonable (again)

As before, we estimated the outcomes using the posterior distributions of the model parameters $p(y^{\mathrm{rep}}|y) = \int p(y^{\mathrm{rep}}|\theta)\, p(\theta|y)\mathrm{d}\theta$. The estimated values by the posterior means (in red) is tighter

---

[3] Multilevel modeling does not move individual group means toward the grand mean; instead, it "moves the error terms toward zero" or "moves the local averages toward their predictions from the regression model" (https://statmodeling.stat.columbia.edu/2023/03/27/48898/). So, when we are predicting well productions given block-level or time-level predictors, multilevel modeling partially pulls the individual block or time means toward the fitted model.

than the observed values (in blue). So, the current model could still be improved. Note there are two predicted values less than 0, which may be set to 0.

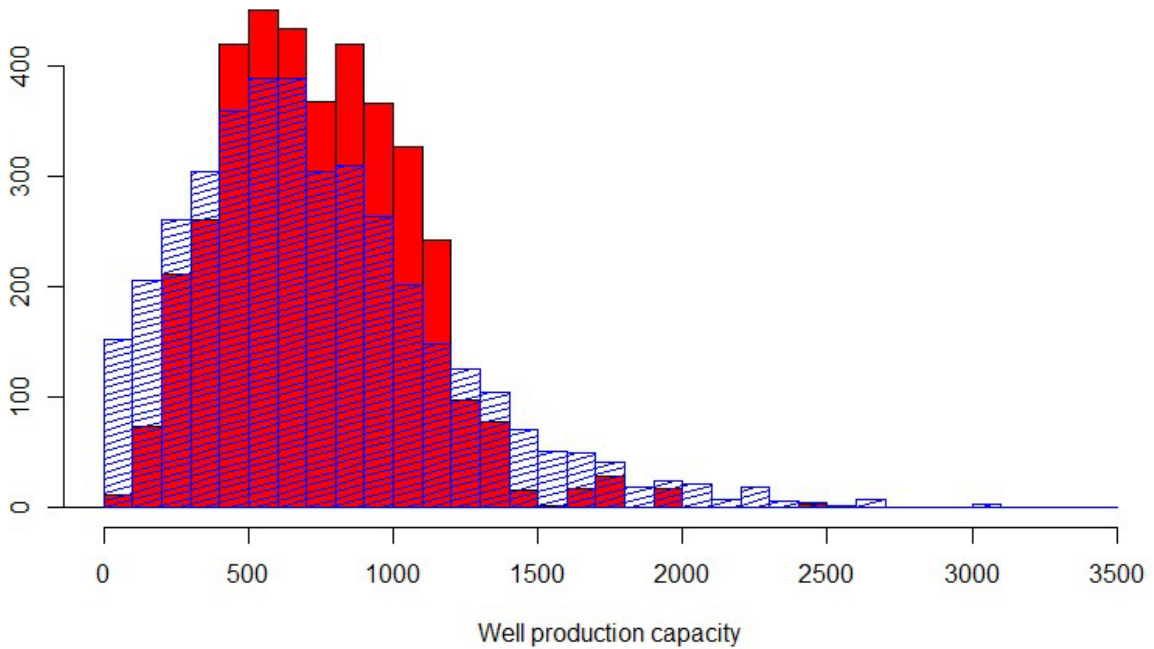

Histograms of well production capacity: Predicted in red vs. data in blue

Computing the root-mean-square deviation (RMSD) between the estimated and the observed, we get about 322 (whether or not the 2 negative estimates are set to zeros), compared to 344 by the initial model. About 90 % of the discrepancies fall between -496 and 555.

## Model-based estimates vs. observed averages

Finally, we averaged the estimated well-level production capacities in each block-time combination as an estimate of the block-time-level production capacity. As an illustration, we show the first ten MHB's whose first four characters are DN87. (The block-time combinations with no observations show NA's.)

|        | Year |      |      |      |      |      |      |      |      |      |
|--------|------|------|------|------|------|------|------|------|------|------|
| MHB    | 2015 | 2016 | 2017 | 2018 | 2019 | 2020 | 2021 | 2022 | 2023 | 2024 |
| DN87au | 234  | NA   | NA   | NA   | NA   | NA   | NA   | NA   | NA   | NA   |
| DN87cm | 343  | NA   | NA   | NA   | NA   | NA   | NA   | NA   | NA   | NA   |
| DN87cq | 357  | NA   | NA   | NA   | NA   | NA   | NA   | NA   | NA   | NA   |
| DN87cw | 308  | NA   | NA   | NA   | NA   | NA   | NA   | NA   | NA   | NA   |
| DN87dd | NA   | NA   | 460  | NA   | NA   | NA   | NA   | NA   | NA   | NA   |
| DN87df | 189  | NA   | NA   | NA   | NA   | NA   | NA   | NA   | NA   | NA   |
| DN87dq | 243  | NA   | NA   | NA   | NA   | NA   | NA   | NA   | NA   | NA   |
| DN87dw | 234  | NA   | NA   | NA   | NA   | NA   | NA   | NA   | NA   | NA   |
| DN87dx | 140  | 239  | NA   | NA   | NA   | NA   | NA   | NA   | NA   | NA   |
| DN87ef | NA   | NA   | NA   | 360  | NA   | NA   | NA   | NA   | NA   | NA   |

The averages of the observed values are also computed as comparisons.

| MHB | Year | | | | | | | | | |
|---|---|---|---|---|---|---|---|---|---|---|
| | 2015 | 2016 | 2017 | 2018 | 2019 | 2020 | 2021 | 2022 | 2023 | 2024 |
| DN87au | 222 | NA | NA | NA | NA | NA | NA | NA | NA | NA |
| DN87cm | 347 | NA | NA | NA | NA | NA | NA | NA | NA | NA |
| DN87cq | 354 | NA | NA | NA | NA | NA | NA | NA | NA | NA |
| DN87cw | 295 | NA | NA | NA | NA | NA | NA | NA | NA | NA |
| DN87dd | NA | NA | 378 | NA | NA | NA | NA | NA | NA | NA |
| DN87df | 91 | NA | NA | NA | NA | NA | NA | NA | NA | NA |
| DN87dq | 210 | NA | NA | NA | NA | NA | NA | NA | NA | NA |
| DN87dw | 147 | NA | NA | NA | NA | NA | NA | NA | NA | NA |
| DN87dx | 217 | 132 | NA | NA | NA | NA | NA | NA | NA | NA |
| DN87ef | NA | NA | NA | 273 | NA | NA | NA | NA | NA | NA |

We believe the model-based estimates are more informative than the simple empirical estimates, because we extract more information from the data through partial pooling by MHB and Time (small area-time estimation), and we regularize the inferences by the priors (Bayesian modeling).

With these estimated values, one can make further inferences, for example, by interpolating between two consecutively observed times or extrapolating for unobserved future times, using some forecasting models. Some of the MHB's could be grouped together. For example, the mean or expected production capacity of $Y$ in MHB 1 and MHB 2 at Time 1 may be modeled by:

$$\mu_Y(b \in \{1,2\}, t = 1) = \frac{\mu_Y(b = 1, t = 1)N_{1,1} + \mu_Y(b = 2, t = 1)N_{2,1}}{N_{1,1} + N_{2,1}},$$

where $N_{1,1}$ and $N_{2,1}$ are the predicted or hypothesized numbers of wells in MHB = 1 and MHB = 2, respectively, at Time = 1. Without additional geographical or geological information, $N_{1,1}$ and $N_{2,1}$ may be assumed to be the same, as MHB's have the same size by construction. Then, it would reduce to an unweighted mean and could be estimated by:

$$\frac{\widehat{\mu_Y}(b = 1, t = 1) + \widehat{\mu_Y}(b = 2, t = 1)}{2}.$$

## 3. Further expansion of the model

The model (5) can be further expanded even without any additional variables. In fact, such an expansion may be necessary for fitting data from different regions or time frames. We introduce the following model expansion:

$$Y_i \sim \text{normal}\big(\alpha_{b[i]} + \tau_{t[i]} + (\gamma_{b[i]}\overline{EW}_{b[i]} + \delta_{b[i]}\overline{ES}_{b[i]} + \varphi_{t[i]}\overline{EW}_{t[i]} + \omega_{t[i]}\overline{ES}_{t[i]})L_i, \quad \sigma_Y\big),$$

$$i = 1, \ldots, N. \quad (7)$$

The expansion is in the slope term for $L$. It separates the water and sand intensities for each $i$:

$$EW_i = \frac{W_i}{L_i} \text{ and } ES_i = \frac{S_i}{L_i}$$

and incorporates their averages for each time $t$ as well as for each MHB $b$:

$$\overline{EW}_b = \frac{\sum_{i \in b} EW_i}{N_b} \text{ and } \overline{ES}_b = \frac{\sum_{i \in b} ES_i}{N_b},$$

where $N_b$ is the number of wells in MHB $b$, and

$$\overline{EW}_t = \frac{\sum_{i \in t} EW_i}{N_t} \text{ and } \overline{ES}_t = \frac{\sum_{i \in t} EW_i}{N_t},$$

where $N_t$ is the number of wells in time $t$.

Recall that the right-skewness in the marginal distributions of $Y_i$ (oil production), $W_i$ (water usage), $S_i$ (sand usage), and $L_i$ (lateral length) was ignored or ignorable in fitting the earlier models. However, it could cause a fitting problem, if it is not modeled. In order to make our approach more robust against the data from other regions or time frames, we now log-transform the variables.

Before the log-transformation, however, any zero values in each variable are replaced by the block-level averages of the variable. A possible justification is that those zero values are not true zeros but represent some data problems in the original data sources, e.g., missing values. (Some records in the current dataset contained zero oil production values.) An MHB block can have only one observation and its value may be zero. Thus, for the replacement or imputation, we use MHB blocks defined by the first four characters of their identification codes, instead of the full six characters.

Meanwhile, we currently accept all non-zero values as they are, though some of them could have measurement errors or biases. If we have good information about them, we would use it to correct them in the data preprocessing stage or to model them in the modeling stage.

In constructing $EW$ and $ES$ from $\log(L)$, $L$ cannot take 1 as $\log(1) = 0$. If that were the case, we would set $EW_i = 0$ and $ES_i = 0$. (In the current data, the minimum lateral length is greater than 1.)

With $EW_i$ and $ES_i$ being computed from the zero-value-imputed and log-transformed $W_i$, $S_i$, and $L_i$, we then standardize $EW_i$ and $ES_i$ as well as $Y_i$ and $L_i$ (not within $EW_i$ or $ES_i$) so that each has the mean 0 and the standard deviation 0.5.

We interpret that $\overline{EW}$ and $\overline{ES}$ independently contribute to the (quad-)linear approximation of the levels of technology or innovation associated with the water usage efficiency level and the sand usage efficiency level, respectively, and each contribution is allowed to vary by block and time.

We continue to use the "weakly informative" priors to stabilize our inferences:

$$\alpha_b \sim \text{normal}(0, 0.5), \quad b = 1, \dots, B,$$

$$\gamma_b \sim \text{normal}(0, 0.5), \quad b = 1 \dots, B,$$

$$\delta_b \sim \text{normal}(0, 0.5), \quad b = 1 \dots, B,$$

$$\tau_1 \sim \text{normal}(0, 0.5), \quad \tau_t \sim \text{normal}(\tau_{t-1}, 0.5), \quad t = 2 \dots, T,$$

$$\varphi_1 \sim \text{normal}(0, 0.5), \quad \varphi_t \sim \text{normal}(\gamma_{t-1}, 0.5), \quad t = 2 \dots, T,$$

$$\omega_1 \sim \text{normal}(0, 0.5), \qquad \omega_t \sim \text{normal}(\gamma_{t-1}, 0.5), \qquad t = 2 \dots, T,$$

and

$$\sigma_Y \sim \text{normal}^+(0, 0.5).$$

This expanded model resulted in a fast and good posterior convergence and produced robust posterior predictions.

The posterior means of the time parameters $\tau_t$ (the intercept), $\varphi_t$ (water usage), and $\omega_t$ (sand usage) are plotted from 2015 to 2024 below. Some possible interpretations of those estimates may be as follows. The base effect of time on the oil production capacity, $\tau_t$, gradually increased to peak in 2021 and plateaued or began to decrease from there. The water usage effect on the oil production capacity, $\varphi_t$, peaked in 2018 and rapidly decreased since 2020, while the sand usage effect on the oil production capacity, $\omega_t$, increased over the time period.

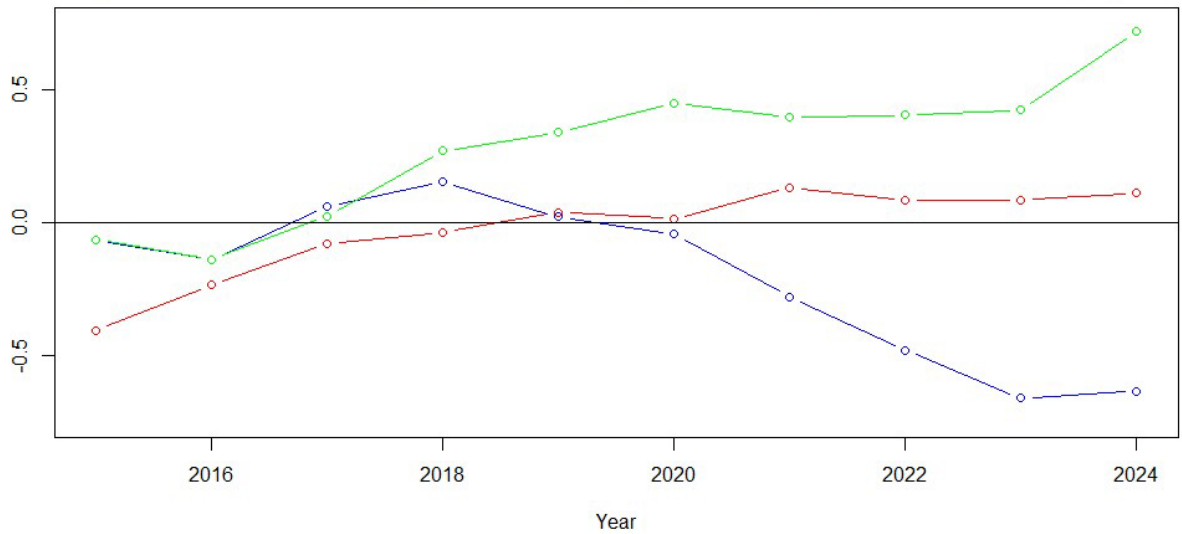

Finally, the model's posterior predictions are shown below.

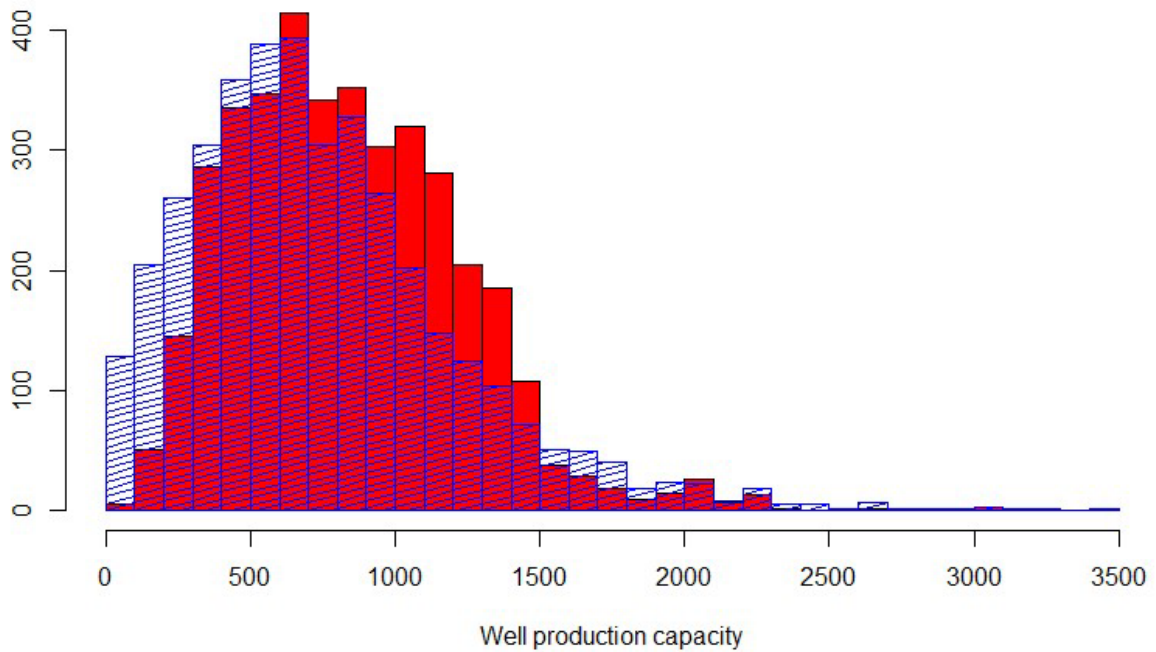

Histograms of well production capacity: Predicted in red vs. data in blue

The RMSD between the estimated and the observed has increased to 346 (no negative estimates here), compared to 322 by the previous model. The current model does a better job in fitting the larger data values while a worse job in fitting the smaller data values, as expected because of the log-transformations—about 90 % of the discrepancies fall in between -605 and 482, compared to between -496 and 555. Practically, however, the current model's performances are not better or worse than those of the previous models at least in these metrics.

Meanwhile, the final model is substantively more complicated than the first two models, i.e., it explains more. That is important in science. Gelman (2004) quoted Neal (1996, pp. 103-104):

> Sometimes a simple model will outperform a more complex model . . . Nevertheless, I believe that deliberately limiting the complexity of the model is not fruitful when the problem is evidently complex. Instead, if a simple model is found that outperforms some particular complex model, the appropriate response is to define a different complex model that captures whatever aspect of the problem led to the simple model performing well.

## 4. Epilogue

Given the continuous predictors $L$, $S$, and $W$ and the categorical predictors $MHB$ and $YEAR$, we have explored how one might model the continuous outcome(s) $Y$. A naïve approach might begin with a model such as:

$$Y_i = \alpha + \beta_1 L_i + \beta_2 S_i + \beta_3 W_i + \varepsilon_i,$$

in which $MHB$ and $YEAR$ may be used to adjust the intercept or/and slopes. This model is simple, but the simplicity comes from strong assumptions—linearity, additivity, homogenous errors, and normal errors. Those assumptions may or may not be reasonable with the particular data one may have, but they tend to be blindly made for convenience. The convenience becomes necessary when the least squares or maximum likelihood estimation is used, particularly if the data are limited in size or variation.

The opposite end of the model-complexity spectrum in terms of parameterization would be:

$$Y_i = \alpha_i + \beta_{1i}L_i + \beta_{2i}S_i + \beta_{3i}W_i + \varepsilon_i.$$

This model is more general than the first one because the intercept and slope are specified for each $i$ in the data. In fact, the least squares or maximum likelihood estimation would practically fail. Even with a Bayesian approach, one would need strong priors to make any sensible posterior inferences.

If we characterize the first formulation as complete pooling, the second one is no pooling. What we can consider is somewhere between, i.e., partial pooling, and the categorical predictors $MHB$ and $YEAR$ become useful.

Using only $L_i$ and changing the notation from $\beta_1$ to $\beta$, we can rewrite the model with the most general grouping of the cases by the two categorical predictors as:

$$Y_i = \alpha_{(b,t)[i]} + \beta_{(b,t)[i]}L_i + \varepsilon_i,$$

where $\alpha_{(b,t)[i]}$ and $\beta_{(b,t)[i]}$ are, respectively, the intercept and slope for the group identified by $MHB = b$ and $YEAR = t$, to which the case $i$ belongs.

Reducing the generality, if we assume independence of the effects by $MHB$ and $YEAR$ on the intercepts and slopes, we can derive the general form of the models considered in this paper:

$$Y_i = \alpha_{b[i]} + \alpha_{t[i]} + (\beta_{b[i]} + \beta_{t[i]})L_i + \varepsilon_i.$$

(Note: Here, the parameters are distinguished by different Greek letter and different indexing.)

We conceptualized the physical relationship among $Y$, $L$, $S$, and $W$ so that $L$, $S$, and $W$ affect $\beta_{b[i]}$ and $\beta_{t[i]}$. Since $\beta_{b[i]}$ and $\beta_{t[i]}$ are group-level parameters, we first created the group-level variables:

$$\overline{EW}_b = \frac{\sum_{i \in b} EW_i}{N_b}, \overline{ES}_b = \frac{\sum_{i \in b} ES_i}{N_b}, \overline{EW}_t = \frac{\sum_{i \in t} EW_i}{N_t}, \text{ and } \overline{ES}_t = \frac{\sum_{i \in t} EW_i}{N_t},$$

where $EW_i = W_i/L_i$, $ES_i = S_i/L_i$, $N_b$ is the number of cases in $MHB = b$, and $N_t$ is the number of cases in $YEAR = t$. Using $\gamma_0, \gamma_1,$ and $\gamma_2$ here, we specified:

$$\beta_{b[i]} = \gamma_{0b[i]} + \gamma_{1b[i]}\overline{EW}_{b[i]} + \gamma_{2b[i]}\overline{ES}_{b[i]}$$

and

$$\beta_{t[i]} = \gamma_{0t[i]} + \gamma_{1t[i]}\overline{EW}_{t[i]} + \gamma_{2t[i]}\overline{ES}_{t[i]}.$$

For the final model in Section 3, we set: $\gamma_{0b[i]} = 0$ and $\gamma_{0t[i]} = 0$. For the spatio-temporal model in Section 2, we set: $\gamma_{0b[i]} = 0$, $\gamma_{1t[i]} = 0$, $\gamma_{2t[i]} = 0$, and $\gamma_{1b[i]} = \gamma_{2b[i]}$. And, for the initial spatial model in Section 1, we did not use the $YEAR$ variable so there were no $YEAR$ parameters $\alpha_t$ or $\beta_t$, and we set: $\gamma_{0b[i]} = \gamma_0$ and $\gamma_{1b[i]} = \gamma_1$ for each $b[i]$ for all $i$ and $\overline{W}_{b[i]} = \frac{\sum_{i \in b} W_i}{N_b}$ (with $W_i$ rather than $EW_i$).

There are different "forking paths" (in a positive sense) that could serve different data or analytic purposes. The paths this paper took are only a few examples of such paths. The Bayesian multilevel modeling framework, however, helped us explore not only effectively and efficiently but also in principled ways.

## 5. References

Gelman, A. (2004, December 10). "Against parsimony." Statistical Modeling, Causal Inference, and Social Science. https://statmodeling.stat.columbia.edu/2004/12/10/against_parsimo/.

Gelman, A. (2006). "Multilevel (Hierarchical) Modeling: What It Can and Cannot Do." Technometrics, 48: 3.

Gelman, A. (2008). "Scaling regression inputs by dividing by two standard deviations." Statistics in Medicine, 27:2865–2873.

Gelman, A. (2020). "Prior Choice Recommendations." https://github.com/stan-dev/stan/wiki/Prior-Choice-Recommendations.

Gelman, A., Hill, J., and Vehtari, A. (2020). Regression and Other Stories. https://users.aalto.fi/~ave/ROS.pdf.

Lieskovsky, J. and Zyren, J. (2022). "Sub-County Analysis of Oil and Gas Wells." Unpublished manuscript.

Neal, R. M. (1996). Bayesian Learning for Neural Networks. https://glizen.com/radfordneal/bnn.book.html.

Stan Development Team (2024). RStan: the R interface to Stan. R package version 2.32.6, https://mc-stan.org/rstan/.

Vehtari, A. (2024). "Cross-validation FAQ." https://mc-stan.org/loo/articles/online-only/faq.html.

Vehtari, A., Gabry, J., Magnusson, M., Yao, Y., Bürkner, P., Paananen, T., Gelman, A. (2024). "loo: Efficient leave-one-out cross-validation and WAIC for Bayesian models." R package version 2.8.0, https://mc-stan.org/loo/.

Vehtari, A., Gelman, A., Simpson, D., Carpenter, B., and Bürkner, P.-C. (2021). "Rank-normalization, folding, and localization: An improved $\hat{R}$ for assessing convergence of MCMC." arXiv preprint arXiv:1903.08008 [stat.CO].